\documentclass[
 amsmath,amssymb,
 prd,
 notitlepage,
 twocolumn
]{revtex4-1}
\usepackage{graphicx}
\usepackage{float}
\usepackage{epstopdf,cancel}
\usepackage{epsf,latexsym,bbm,euscript}
\usepackage{amssymb,amsmath}
\usepackage{mathtools} 
\usepackage{times,graphics}
\usepackage{soul,xcolor}
\usepackage{mathtools}
\usepackage{mathrsfs}

\usepackage{natbib}

\usepackage{bm}

\def\6{{\langle}}
\def\9{{\rangle}}
\newcommand{\defeq}{\vcentcolon=}
\newcommand{\eqdef}{=\vcentcolon}

\def\etal{\textit{et al.}}

\usepackage{url,hyperref}
\hypersetup{colorlinks,linkcolor={blue!55!black},citecolor={red!50!black},urlcolor={blue!45!black}}

\begin{document}

\title{Covariant formulation of spin optics for electromagnetic waves}

\author{Pravin Kumar Dahal}
 \email{pravin-kumar.dahal@hdr.mq.edu.au}
\affiliation{School of Mathematical \& Physical Sciences, Macquarie University}

\date{\today}

\begin{abstract}
    We develop geometric optics expansion up to the subleading order for circularly polarized electromagnetic waves on curved spacetime. This subleading order geometric optics expansion, in which the conventional eikonal function is modified by inserting a carefully chosen helicity-dependent correction, is called spin optics. We derive the propagation and polarization equations in the spin optics approximation as electromagnetic waves travel in curved spacetime. Polarization-dependent deviation of the light ray trajectory from the geodesic, describing the gravitational spin Hall effect, is observed. We also establish an analogy with the related phenomena (optical Magnus effect) of condensed matter physics.
\end{abstract}


\maketitle

\section{Introduction}

Geometric optics is valid in the infinitely large frequency limit~\cite[]{mtw, d35, d36}. In this approximation, electromagnetic waves propagating on the fixed curved background follow a null ray trajectory. At large but finite frequencies, the geometric optics approximation no longer remains valid, as backreaction from the helicity may cause ray trajectory to depart from the geodesics by a significant amount. The gravitational spin Hall effect refers to this helicity dependent phenomenon of the propagation of a light ray in curved spacetime in the subleading order~\cite[]{f31,38,39,40}. This is similar to the gravitational Faraday effect (or the spin Hall effect for gravitational waves), which is observed for high-frequency polarized gravitational waves propagating in curved spacetimes~\cite[]{p35,41,42}.

The spin Hall effect is due to the interaction of polarization/spin with the orbital motion of the rays~\cite[]{43,44,45}. Related phenomena are observed in different areas of physics, including optics, where the polarization-dependent deflection of light was predicted~\cite[]{1,2,3} and verified experimentally for both electrons~\cite[]{wks17,kmg18} and light~\cite[]{4,5} (also see Ref.~\cite{aw55} for the explanation of these results in terms of classical optics). When an electromagnetic wave propagates in an inhomogeneous medium, the spin Hall effect of light (also known as the optical Magnus effect) is observed. The coupling of spin with the orbital motion comes from the interaction of the polarization of waves with the refractive index gradient of the medium. As a result, electromagnetic waves are deflected transversally in a direction perpendicular to the refractive index gradient. The spin Hall effect can be expressed in terms of Berry curvature~\cite[]{6,abn54,7} and provides correction to the geometric optics, which is approximately proportional to the frequency inverse, in the subleading order. This phenomenon of the optical Magnus effect can be extended to general relativity, where spacetime curvature itself acts as an inhomogeneous medium~\cite[]{12,13,14,f31,38,46,47}. This analogy comes from the fact that the equations for electromagnetic waves propagating in some optical medium and curved spacetime are identical. Thus, the gravitational spin Hall effect is caused by the interaction of polarisation with spacetime curvature, requiring a spin-dependent correction to particle dynamics.

Various approaches can be found in the literature for the calculation of the gravitational spin Hall effect, and they can be broadly classified into three categories: 1) use of the Souriau-Saturnini equations~\cite[]{10,11}, which is a modification of the Mathisson-Papapertou-Dixon equation~\cite[]{8,9,48,49} for the motion of massive spinning particles to the massless ones, 2) application of the methods of quantum mechanics, such as using the Foldy-Wouthyusyen transformation on the Bargmann-Wigner equations~\cite[]{b24,47,50}, and 3) spin optics~\cite[]{f31,d35,d36,39}. See the review by~\cite{12} and the references therein for the detailed discussions of each of these approaches and their comparisons. Here, we focus on spin optics (or modified geometric optics), where the eikonal function is modified by including the spin-dependent term. This correction is of the order $\omega^{-1}$ and thus suppressed in the high-frequency approximation. However, such correction becomes essential for the propagation of polarized rays of high but finite frequencies at large distances (where the ray trajectory could be modified).

Self-dual and anti-self-dual solutions of the Maxwell equations represent electromagnetic fields with right- and left-hand circular polarization, respectively. The use of the Wentzel-Kramers-Brillouin (WKB) ansatz for each of these solutions does not require the same eikonal functions for both types of waves. We instead include the helicity-dependent first-order high-frequency correction in the eikonal function in the spin optics approximation. The development of spin optics for general spacetime has been demonstrated in recent papers by Refs.~\cite{13} and ~\cite{14}. Their results are slightly different from each other and our paper. We will compare these results.

Here, we formulate the theory of spin optics and develop self-consistent ray and transport equations for the propagation of electromagnetic waves in curved spacetime. Electromagnetic waves from astrophysical sources might propagate over cosmological distances before reaching the observer. As a result, they might encounter inhomogeneities in the form of spacetime curvature. These inhomogeneities act as a lensing object, and if their length scale is much larger than the characteristic wavelength of the electromagnetic waves, then the geometric optics approximation is valid. However, there might be a situation, where the characteristic wavelength is small but not negligible compared to the length scale of the variation of spacetime curvature. In such a situation, the necessity of the subleading order correction to the geometric optics arises while studying the gravitational lensing of electromagnetic waves. The standard lensing phenomena likely include wave effects, changing the propagation properties of waves~\cite{t26}. In Sec.~\ref{s2}, we write the Maxwell equations in curved spacetime and obtain the equation for electromagnetic waves. To solve the electromagnetic wave equation, we present the WKB ansatz for the vector potential in Sec.~\ref{s3}. Next, we show that we can reduce the geometric optics solution to a set of Fermi transported null tetrads and then generalize this (requirement) to the subleading order. We impose more constraints on the subleading order equations considering the solutions to be self-dual and anti-self-dual. These requirements and considerations are sufficient to obtain the propagation and polarization equations up to the subleading order. We do this in Sec.~\ref{s4}. We also obtain the stress-energy tensor of the electromagnetic field up to the subleading order approximation in Sec.~\ref{s5} to see the energy flow as the wave propagates. We then compare our results with the literature using the WKB approximation, the same formalism as used here, in Sec.~\ref{s6}. Finally, we discuss our results in Sec.~\ref{s7}.

In this article, we consider the metric $g_{\mu\nu}$ of signature $\left(-,+,+,+\right)$ in a Lorentzian manifold $M$. The phase space is the cotangent bundle $T^* M$ and its points are written as $(x, p)$. Similarly, we write $\tilde z$ for the complex conjugate of $z$ and adopt the Einstein summation convention. We use the system of units with $G=c=1$. A semicolon ($;$) denote the covariant derivative, $\lambda$ denotes the parameter of electromagnetic wave curves, and $\dot x= d x/d\lambda$. We use the sign convention for the curvature adopted in ~\cite{mtw}.

\section{Maxwell equations in curved spacetimes} \label{s2}

In flat spacetime, the Maxwell equations of electrodynamics are
\begin{align}
    &\frac{\partial}{\partial x^\alpha}F^{\alpha\beta}=-J^\beta,\\
    & \frac{\partial}{\partial x^\alpha}F^{\beta\gamma}+\frac{\partial}{\partial x^\beta}F^{\gamma\alpha}+\frac{\partial}{\partial x^\gamma}F^{\alpha\beta}=0,
\end{align}
where $F^{\alpha\beta}$ is the antisymmetric field tensor with components
\begin{equation}
    F^{jk}=e^{jkl} B_l, \qquad F^{0j}=E_j.
\end{equation}
Here, $e^{jkl}$ is the Levi-Civita symbol in three dimensions, and $J^\alpha$ is the conserved four-current. The principle of general covariance implies that these equations hold in a general curved spacetime if covariant derivatives replace the partial derivatives occurring in the equations. Thus, the Maxwell equations in curved spacetime are
\begin{align}
    & F^{\alpha\beta}_{~~~~;\alpha}=-J^\beta,\label{me4}\\
    & F_{\alpha\beta;\gamma}+F_{\gamma\alpha;\beta}+F_{\beta\gamma;\alpha}=0. \label{me5}
\end{align}
Eq.\eqref{me5} allows us to represent the electromagnetic field tensor $F^{\alpha\beta}$ in terms of the vector potential $A^\alpha$ as
\begin{equation}
    F_{\alpha\beta}=A_{\beta;\alpha}-A_{\alpha;\beta}.
\end{equation}
We can use available gauge freedom in the Maxwell equations such that the vector $A^\alpha$ satisfies the Lorenz gauge condition, $A^\alpha_{;\alpha}=0$. We substitute this into Eq.~\eqref{me4} to obtain the wave equation
\begin{equation}
    -A^{\alpha;\beta}_{~~~~;\beta}+R^\alpha_\beta A^\beta=J^\alpha,\label{we7}
\end{equation}
where $R^\alpha_\beta$ denotes the Ricci tensor.

\section{WKB approximation} \label{s3}

Spin optics approximation is valid when the typical wavelength of the wave is very small (but nonnegligible) in comparison with the length scale over which its amplitude and wavelength vary and the radius of curvature of the spacetime on which it propagates. We can locally approximate the wave as a ray propagating on approximately flat spacetime in such a case. Mathematically, we formulate the spin optics approximation using the WKB ansatz as
\begin{equation}
    A^\alpha=a^\alpha e^{i \omega {\cal S}},\label{a8}
\end{equation}
where $a^\alpha$ is the complex amplitude that varies slowly, and $\omega {\cal S}$ is the real phase that changes rapidly. Here, $\omega$ is the characteristic frequency of the problem. The wave vector is the phase gradient, that is, $l_\alpha={\cal S}_{;\alpha}$. Let us write the square amplitude $a=\left(\tilde a^\alpha a_\alpha\right)^{1/2}$ and the polarization vector $m^\alpha=a^\alpha/a$. We can expand the wave vector and polarization vector in powers of $1/\omega$ as
\begin{align}
    l^\alpha=& l^\alpha_0+\frac{l^\alpha_1}{\omega}+\frac{l^\alpha_2}{\omega^2}+ ...,\\
    m^\alpha=& m^\alpha_0+\frac{m^\alpha_1}{\omega}+\frac{m^\alpha_2}{\omega^2}+ ....
\end{align}
We should mention here that we cannot absorb higher order phase factors like ${\cal S}_1(\lambda)$ into the complex amplitude $m^\alpha_0$ by transformation $m^\alpha \to e^{i {\cal S}_1 (\lambda)/\omega} m^\alpha$. This is because we will use the Fermi propagated null tetrad in Sec.~\ref{ntb} and two of its components, $l^\alpha$ and $m^\alpha$, give the trajectory and polarization of the waves, respectively. Fermi propagation reduces the freedom in the transformation of the null tetrad $m^\alpha \to e^{i {\cal S}_1 (\lambda)/\omega} m^\alpha$ by the condition
\begin{equation}
    \frac{d {\cal S}_1 (\lambda)}{d\lambda}= 0. \label{gt11}
\end{equation}
This is the reason why we should expand both the wave vector and polarization vector separately in powers of $\omega$. Let us substitute this vector potential onto the source free wave equation (Eq.~\eqref{we7} with $J^\alpha=0$) and the Lorenz gauge condition. We start from the Lorenz condition, which up to the subleading order in $\omega$ can be written as
\begin{equation}
    l_0^\alpha m_{0 \alpha}+\frac{1}{\omega}\left(l_0^\alpha m_{1 \alpha}+l_1^\alpha m_{0 \alpha}- i \left(\frac{a_{;\alpha}}{a}m_0^\alpha+m^\alpha_{0;\alpha}\right)\right)=0.\label{lc11}
\end{equation}
Again, we substitute the vector potential into the source-free wave equation, which gives
\begin{multline}
    j^\alpha \eqdef m_0^\alpha l_{0\beta} l_0^\beta+\frac{1}{\omega}\bigg(m_1^\alpha l_{0\beta} l_0^\beta+ 2 m_0^\alpha l_{1\beta} l_0^\beta\\
    -i \left(m_0^\alpha l^\beta_{0;\beta}+2 m^\alpha_{0;\beta}l_0^\beta+2 \frac{a_{;\beta}} {a}m_0^\alpha l_0^\beta \right) \bigg)=0,\label{we12}
\end{multline}
up to the subleading order in $\omega$. We now calculate $\tilde m_{0\alpha}j^\alpha+ m_{0\alpha} \tilde j^\alpha$, which is the identically vanishing quantity
\begin{equation}
     l_{0\beta} l_0^\beta+\frac{2}{\omega} \left(l_{1 \beta}-b_\beta \right) l_0^\beta =0.\label{dr13}
\end{equation}
This is the generalization of the dispersion relation of geometric optics. Here, we have used $\tilde m_0^\alpha m_{0 \alpha}=1$ and substituted
\begin{equation}
    \frac{i}{2}\left(\tilde m^\alpha m_{\alpha;\beta}- m^\alpha \tilde m_{\alpha; \beta} \right)= i \tilde m^\alpha m_{\alpha;\beta} \defeq b_\beta. \label{b14}
\end{equation}

\section{Spin optics approximation} \label{s4}

\subsection{Propagation and polarization vector as tetrad components}\label{ntb}

Let us construct a set of null tetrads $\left(l_0^\alpha, n_0^\alpha, m_0^\alpha, \tilde m_0^\alpha\right)$ satisfying the following orthogonality and completeness relationships
\begin{align}
    & l_0^\alpha m_{0 \alpha}= l_0^\alpha l_{0\alpha}= l_0^\alpha \tilde m_{0\alpha}=0,  \qquad m_0^\alpha \tilde m_{0 \alpha}=1, \label{oc18}\\
    & m_0^\alpha m_{0 \alpha}= \tilde m_0^\alpha \tilde m_{0\alpha}=0,\label{oc19}\\
    & n_0^\alpha m_{0 \alpha}= n_0^\alpha n_{0\alpha}= n_0^\alpha \tilde m_{0\alpha}=0, \qquad n_0^\alpha l_{0\alpha}=-1. \label{oc20}
\end{align}
Comparison of Eqs.~\eqref{oc18} with the equations of geometric optics (see Eqs.~\eqref{go15}) shows that two components of the tetrad $l_0^\alpha$ and $m_0^\alpha$ could be identified with the wave vector and polarization vector, respectively. Auxiliary null vectors $n_{0 \alpha}$ and $m_{0 \alpha}$ are not unique and can be chosen in such a way that they satisfy Eqs.~\eqref{oc19} and \eqref{oc20}. We will discuss below in Sec.~\ref{pb} that the validity of Eqs.~\eqref{oc19} determines the state of polarization; specifically, circularly polarized waves satisfy these relations. Moreover, these components of the null tetrad also satisfy
\begin{align}
    & l^\alpha_{0;\beta}l_0^\beta=0, \qquad m^\alpha_{0;\beta}l_0^\beta=0=\tilde m^\alpha_{0;\beta}l_0^\beta, \label{pt21}\\
    & n^\alpha_{0;\beta}l_0^\beta=0 ,\label{pt22}
\end{align}
where Eqs.~\eqref{pt21} again follows from geometric optics (to obtain the first relation, we have applied $l_{0 \alpha;\beta}=l_{0 \beta;\alpha}$). We can choose $n_0^\alpha$, such that it satisfies Eq.~\eqref{pt22}. We can show that this choice is indeed possible by introducing the Fermi derivative operator ${\cal D}_l$ along the ray $l^\alpha$, which applied to the vector $A^\alpha$ gives~\cite[]{f31}
\begin{equation}
    {\cal D}_l A^\alpha =l_0^\gamma A^\alpha_{;\gamma}-w_\gamma A^\gamma n^\alpha+ A^ \gamma n_{ \gamma} w^\alpha,
\end{equation}
where $w^\alpha =l_0^\gamma l^\alpha_{;\gamma}$ vanishes identically in geometric optics. We have ${\cal D}_l l^\alpha=0$ as $l^\gamma l_\gamma=0$. If the Fermi derivative ${\cal D}_l A^\alpha$ of a vector $A^\alpha$ is zero, then it is said to be Fermi propagated, and we can easily prove that the scalar product of any two Fermi propagated vectors is constant. We can apply this to the set of tetrads $\left(l^\alpha, n^\alpha, m^\alpha, \tilde m^\alpha\right)$: they satisfy the orthogonality and completeness relations similar to the one given in Eqs.~\eqref{oc18}-\eqref{oc20} everywhere on the ray if they satisfy those relations the at some point on the ray and if they all are Fermi propagated. As Fermi propagation preserves the scalar product, the set of null tetrad $\left(l^\alpha, n^\alpha, m^\alpha, \tilde m^\alpha\right)$ satisfies the orthogonality and completeness relations like those in Eqs.~\eqref{oc18}-\eqref{oc20} along the ray and obeys
\begin{align}
    & l_0^\beta n^\alpha_{;\beta}= w^\beta n_{ \beta} n^\alpha,\\
    & l_0^\beta m^\alpha_{;\beta}= w^\beta m_{ \beta} n^\alpha,\\
    & l_0^\beta \tilde m^\alpha_{;\beta}= w^\beta \tilde m_{ \beta} n^\alpha.
\end{align}
Next, let us take advantage of flexibility in selecting a null tetrad
\begin{equation}
    l^\alpha \to F l^\alpha, n^\alpha \to F^{-1} n^\alpha,
\end{equation}
to fix $w^\beta n_{ \beta}=0$, where $F$ is some real function. This condition defines the parameter $\lambda$ along the ray up to some rescaling $\lambda \to F^{-1}\lambda$, and such a choice is called a canonical parametrization~\cite[]{14}. Therefore, in the canonical parametrization, we have
\begin{equation}
    l_0^\beta n^\alpha_{;\beta}= 0, \qquad
    l_0^\beta m^\alpha_{;\beta}= w^\beta m_{ \beta} n^\alpha,\qquad
    l_0^\beta \tilde m^\alpha_{;\beta}= w^\beta \tilde m_{ \beta} n^\alpha.\label{pe28}
\end{equation}
These relations generalize Eqs.~\eqref{oc18}-\eqref{pt22} of geometric optics.

\subsection{Self-dual and anti-self-dual solutions of Maxwell equations}\label{pb}

One can define the complex version of the electromagnetic field tensor $F^{\alpha\beta}$ as
\begin{equation}
    {\cal F}^s=F+ i s F^*,
\end{equation}
where $s= \pm 1$ and $F^*=\epsilon_{\alpha\beta\mu\nu} F^{\mu\nu} /2$ is the Hodge dual of $F^{\alpha\beta}$. Here, $\epsilon_{\alpha\beta\mu\nu}$ is the Levi-Civita symbol in four dimensions, and its components in the tetrad basis are $i l \wedge n \wedge m \wedge \tilde m$.
As $(F^*)^*=-F$, we have $\left({\cal F}^s\right)^*= -i s {\cal F}^s$, a feature due to which we call ${\cal F}_{\alpha\beta}^{s}$ self- (or anti-self-) dual antisymmetric field for $s=+1 (\mathrm{or} -1)$. A general self-dual antisymmetric field can be expanded in terms of the self-dual basis
\begin{equation}
    \left(\mathbf U, \mathbf V, \mathbf W\right) =\left( \tilde m \wedge n, l \wedge m, m \wedge \tilde m- l \wedge n \right),
\end{equation}
as
\begin{equation}
    {\cal F}^{+1}= \Phi_0 \mathbf{U}+ \Phi_1 \mathbf{W}+ \Phi_2\mathbf{V}. \label{pb31}
\end{equation}
In the limit of geometric optics, $\Phi_0=\Phi_2=0$. Now, the field ${\cal F}^{+1}_{\alpha\beta}$ corresponding to the potential of Eq.~\eqref{a8} is
\begin{equation}
    {\cal F}^{+1}_{\alpha\beta}= i \omega {\cal Z}_{\alpha\beta} e^{i {\cal S}},\label{ft32}
\end{equation}
where
\begin{equation}
    {\cal Z}_{\alpha\beta}=l_\alpha a_\beta-l_\beta a_\alpha-\frac{i} {\omega}\left(a_{\beta;\alpha}- a_{\alpha;\beta}\right).\label{fa33}
\end{equation}
Using the condition that the contraction of the self-dual field with the anti-self-dual field vanishes, we obtain
\begin{align}
    & {\cal Z}_{\alpha\beta} m^\alpha n^\beta=0,\label{pe34}\\
    & {\cal Z}_{\alpha\beta} \left(\tilde m^\alpha m^\beta- l^\alpha n^\beta \right)=0,\label{pe35}\\
    & {\cal Z}_{\alpha\beta} l^\alpha \tilde m^\beta=0. \label{pe36}
\end{align}
An anti-self-dual solution can be found by the complex conjugation of the amplitude ${\cal Z}_{\alpha\beta}$ of the self-dual field. Eq.~\eqref{pe36} is satisfied identically in the geometric optics approximation, which can be verified easily by substituting the value of ${\cal Z}_{\alpha\beta}$ from Eq.~\eqref{fa33}. However, Eqs.~\eqref{pe34} and \eqref{pe35} gives
\begin{equation}
    m_0^\alpha m_{0 \alpha}=0 = l_0^\alpha m_{0\alpha}.
\end{equation}
These relations are presented in Eqs.~\eqref{oc18} and \eqref{oc19} as the orthogonality conditions.

\subsection{Equations of spin optics}\label{soD}

\subsubsection{Generalization of the Hamilton-Jacobi equation}

The dispersion equation of geometric optics (see Eq.~\eqref{go15}) could be written as
\begin{equation}
    \frac{1}{2} g_{\alpha\beta} l_0^\alpha l_0^\beta=0, \label{go51}
\end{equation}
which is the Hamilton-Jacobi equation for the leading order phase function ${\cal S}_0$ defined as ${\cal S}_{0;\alpha}= l_{0\alpha}$. To solve this Hamilton-Jacobi equation, we define a Hamiltonian function on the cotangent bundle $T^* M$ as
\begin{equation}
    H(x, l)= \frac{1}{2}g^{\alpha\beta} l_{0\alpha} l_{0 \beta},\label{ham53}
\end{equation}
where $x^\alpha(\lambda)$ is the integral curve of $l^\alpha$. We obtain the following Hamilton's equations of motion
\begin{equation}
    \frac{d x_0^\mu}{d \lambda}= \frac{\partial H}{\partial l_{0\mu}}= g^{\mu\nu} l_{0 \nu}, \label{he54}
\end{equation}
and
\begin{equation}
    \frac{d l_{0\alpha}}{d \lambda}= \frac{\partial H}{\partial x_0^\alpha}=  -\frac{1}{2} \dot x_0^\mu \dot x_0^\nu \frac{\partial g_{\mu\nu}}{\partial x_0^\alpha}, \label{ge42}
\end{equation}
where Eq.~\eqref{he54} and the relation
\begin{equation}
    \frac{\partial g_{\alpha\beta}}{\partial x_{0\mu}}= -g_{\nu\alpha} g_{\rho\beta}\frac{\partial g^{\nu\rho}}{\partial x_{0\mu}}, \label{id43}
\end{equation}
are used in obtaining this. Given a solution of Hamilton's equations of motion, the corresponding solution of the Hamilton-Jacobi equation~\eqref{go51} is obtained from~\cite{g32}
\begin{equation}
    {\cal S}_0 (x,l)= \int_{\lambda} (\dot x_0^\alpha l_{0\alpha}- H(x,l)) d\lambda. \label{a44}
\end{equation}
The Euler-Lagrange equation for this action, whose Lagrangian is the Legendre transformation of the  Hamiltonian in Eq.~\eqref{ham53}, is the geodesic equation. Therefore, Eq.~\eqref{ge42} describes null geodesics, and further simplification of this equation gives
\begin{equation}
    \frac{d \left(g_{\alpha\beta} \dot x_0^\beta\right)}{d \lambda}- \frac{1}{2} \dot x_0^\mu \dot x_0^\nu \frac{\partial g_{\mu\nu}}{\partial x_{0\alpha}}= \frac{D^2 x_0^\alpha}{D\lambda^2}=0,\label{he57}
\end{equation}
where $D/D\lambda$ denotes the covariant derivative along the curve $x^\alpha(\lambda)$. The two Eqs.~\eqref{he54} and \eqref{he57} obtained by solving Hamilton's equations of motion constitute the two equations of geometric optics. We have thus shown that the Hamiltonian defined in Eq.~\eqref{ham53} correctly reproduces the equations of geometric optics. Substituting Eqs.~\eqref{ham53} and \eqref{he54} into the action of Eq.~\eqref{a44} gives
\begin{equation}
    {\cal S}_0= \frac{1}{2} \int \dot x_0^\alpha \dot x_{0\alpha} d\lambda.
\end{equation}
Now, to evaluate the trajectory equation up to the subleading order, we consider the generalized dispersion Eq.~\eqref{dr13} and write it as
\begin{equation}
    \frac{1}{2} g_{\alpha\beta}l_0^\alpha l_0^\beta+\frac{1}{\omega} g_{\alpha\beta} \left(l_1^ \alpha-b^\alpha \right) l_0^\beta =0, \label{hj47}
\end{equation}
which is the Hamilton-Jacobi equation for the subleading order phase function ${\cal S}$ defined as ${\cal S}_{;\alpha}= l_{0\alpha}+ l_{1\alpha}/\omega$. The corresponding Hamiltonian function on the cotangent bundle $T^* M$ is
\begin{multline}
    H(x,l)= \frac{1}{2} g^{\alpha\beta}l_{0\alpha} l_{0\beta}+\frac{1}{\omega} g^{\alpha\beta} \left(l_{1 \alpha}-b_\alpha \right) l_{0\beta}\\
    = \frac{1}{2\omega^2} g^{\alpha\beta} \left(\omega l_{0 \alpha} + l_{1\alpha}- b_\alpha\right) \left(\omega l_{0\beta} + l_{1\beta}- b_\beta\right).\label{he65}
\end{multline}
Hamilton's equations of motion are
\begin{equation}
    \frac{d x^\alpha}{d \lambda}= \frac{\partial H}{\partial l_{\alpha}}= g^{\alpha\beta} \left(l_ \beta- \frac{b_\beta}{\omega}\right), \label{tr64}
\end{equation}
and
\begin{equation}
    \frac{d l_\alpha}{d \lambda}= -\frac{\partial H}{\partial x^\alpha}=  \frac{1}{2} \dot x^\mu \dot x^\nu \frac{\partial g_{\mu\nu}}{\partial x^\alpha}+ \frac{1}{\omega} g^{\mu\nu} \dot x_\nu \frac{\partial b_\mu}{\partial x^\alpha}, \label{eom50}
\end{equation}
where we have used Eqs.~\eqref{id43} and \eqref{tr64} in obtaining this. Thus, the corresponding solution of the Hamilton-Jacobi equation~\eqref{hj47} is obtained from~\cite{g32}
\begin{multline}
    {\cal S} (x,l)= \int_{\lambda} (\dot x^\alpha l_\alpha- H(x,l)) d\lambda\\
    = \frac{1}{2} \int \dot x^\alpha \dot x_\alpha d\lambda + \frac{1}{\omega} \int b_\alpha \dot x^\alpha d\lambda= {\cal S}_0- {\cal S}_B, \label{a51}
\end{multline}
where Eqs.~\eqref{he65} and \eqref{tr64} are used for simplification. In addition to the scalar phase, photons acquire a polarization-dependent phase ${\cal S}_B$. This is the Berry geometric phase acquired by the circularly polarized modes propagating in curved spacetime~\cite[]{b26,z27,v28}. This form of action was considered by Refs.~\cite{14} and \cite{37} to derive the spin Hall effect of light. The first term is the optical path length and its variation yields
\begin{multline}
    \frac{1}{2} \delta\int \dot x^\mu \dot x_\mu d\lambda= \int \dot x_\mu \frac{D \delta x^\mu}{D\lambda} d\lambda\\
    = - \int \frac{D^2 x_ \mu}{D\lambda^2} \delta x^\mu d\lambda.
\end{multline}
Similarly, the second term resembles the Berry connection of optics (see Sec.~\ref{4c2}), and its variation gives
\begin{multline}
    \frac{1}{\omega} \delta\int b_\alpha \dot x^\alpha d\lambda= \frac{1}{\omega} \int \delta b_\alpha \dot x^\alpha d\lambda+ \frac{1}{\omega} \int b_\alpha \frac{D \delta x^\alpha}{D\lambda} d\lambda\\
    = \frac{1}{\omega} \int b_{\alpha;\beta} \dot x^\alpha \delta x^\beta d\lambda- \frac{1}{\omega} \int b_{\beta;\alpha} \delta x^\beta \dot x^\alpha d\lambda.
\end{multline}
Thus, the variational principle $\delta{\cal S}=0$ gives
\begin{equation}
    \frac{D^2 x_\mu}{D\lambda^2}+ \frac{1}{\omega} \left( b_{\mu;\nu}- b_{\nu;\mu}\right) \dot x^\nu= 0. \label{vp62}
\end{equation}
One can also show that simplification of Eq.~\eqref{eom50} gives this exact same equation~\cite[]{14}, as should be the case. In the subleading order geometric optics approximation, the action acquires a new topological term depending on the wave polarization (because the action for the left-handed circularly polarized waves would be the same except for the replacement $\omega\to -\omega$). This term gives the deflection of the ray trajectory in the transverse direction. The resulting phenomenon is called the spin Hall effect because of the coupling of spin with the curved ray trajectory. As could be seen from Eq.~\eqref{tr64}, the topological term makes velocity and momentum noncollinear, which is also the characteristic of waves travelling in anisotropic media (see, for example, ~\cite{z33}).

\subsubsection{Propagation equation in the subleading order} \label{4c2}

To calculate the subleading order terms in the geometric optics approximation, we take the dispersion relation of Eq.~\eqref{dr13}
\begin{equation}
    \left(l_{0\mu}+\frac{1}{\omega}(l_{1\mu}-b_\mu) \right) \left(l_0^\mu +\frac{1}{\omega}(l_1^\mu-b^\mu) \right)=\dot x_\mu \dot x^\mu=0,\label{dr50}
\end{equation}
where we have used Eq.~\eqref{tr64}. We can see that in the leading order in $1/\omega$, $\dot x^\mu= l_0^\mu$ is the tangent vector. This equation suggests that the electromagnetic wave trajectory is still null in the spin optics approximation. However, it is not a geodesic, as evidenced from the equation
\begin{equation}
    \frac{D\dot x^\mu}{D\lambda}=\frac{1}{\omega} \left(l^\mu_{1;\nu}-b^\mu_{;\nu}\right)l_0^\nu.\label{gd39}
\end{equation}
Comparing this with Eq.~\eqref{vp62}, we get $l_{1 \alpha;\beta}=b_{\beta;\alpha}$. Further simplification of $b_{\beta;\alpha}- b_{\alpha;\beta} \defeq k_{\alpha\beta}$ gives
\begin{equation}
    k_{\alpha\beta}=-i R_{\alpha\beta\mu\nu}m^\mu \tilde m^\nu+ i \left(\tilde m^\nu_{;\alpha} m_{\nu;\beta}- \tilde m^\nu_{;\beta} m_{\nu;\alpha}\right). \label{k58}
\end{equation}
We substitute this back into Eq.~\eqref{gd39} to obtain
\begin{equation}
    \frac{D^2 x^\alpha}{D\lambda^2}=-\frac{i}{\omega} R^\alpha_{~\beta\mu\nu}m^\mu \tilde m^\nu l_0^\beta \approx -\frac{i}{\omega} R^\alpha_{~\beta\mu\nu} l_0^\beta m_0^\mu \tilde m_0^\nu.\label{te43}
\end{equation}
Thus, in the spin optics approximation, light travels in a null but nongeodesic trajectory. Let us interpret this result by comparing it with the related phenomena in condensed matter physics. The Lagrangian corresponding to the action of Eq.~\eqref{a51} could be written as
\begin{multline}
    {\cal L}= {\cal L}_0+ {\cal L}_1; \qquad {\cal L}_0= \frac{1}{2} \dot x^\alpha \dot x_\alpha,\\
    {\cal L}_1= \frac{1}{\omega} b_\alpha \dot x^\alpha= - \frac{1}{\omega} \mathscr{A}_\alpha \dot x^\alpha, \label{l101}
\end{multline}
where ${\cal L}_0$ is the Lagrangian corresponding to the leading order geometric optics and ${\cal L}_1$ gives the spin-orbit coupling. Here, $\mathscr A_\alpha= - b_\alpha= - i \tilde m^\beta m_{\beta;\alpha}$ could be identified with the Berry gauge field (this has the same form as the spin-orbit interaction of light in gradient-index medium and spin-orbit interaction of electrons occurring in Dirac equation, see, for example, Refs.~\onlinecite{37,b22,b24,s25}). The nonrelativistic version of ${\cal L}_1$ appears in the theory of spinning particles in ~\cite{37, b19} (also see, for example, ~\cite{f20, b29}). ${\cal L}_1$ introduces an additional polarization-dependent wave phase, which could be explained as the Berry phase. This Berry phase manifests itself dynamically, thereby inducing an additional term in the equation of motion of the ray trajectory that describes the spin Hall effect. Thus, the Berry phase and spin Hall effect together characterize the spin-orbit interaction  of electromagnetic waves.

Berry connection $\mathscr A_\alpha$ appears in the Lagrangian as an external vector potential affecting the light trajectory. The curvature associated with the Berry connection can be defined as
\begin{equation}
    \frac{\partial \mathscr A_\alpha}{\partial x^\beta}- \frac{\partial \mathscr A_\beta}{\partial x^\alpha}= b_{\beta;\alpha}-b_{\alpha;\beta} \equiv k_{\alpha\beta}.
\end{equation}
This quantity is known by the name of Berry curvature and plays the role of a field strength tensor corresponding to the vector potential $\mathscr A_\alpha$.

\subsubsection{Polarization equation in the subleading order}

The polarization vector depends only on the direction of the ray trajectory (or the momentum of photons). The momentum of free particles propagating in curved spacetime is a function of position only. Therefore, the Berry connection determines the evolution of the wave polarization in curved spacetime. Let us substitute $w^\alpha =l_0^\beta l^\alpha_{;\beta}$ from Eq.~\eqref{gd39} into the Eqs.~\eqref{pe28} to obtain equations for the evolution of the polarization vector
\begin{align}
    l_0^\beta n^\mu_{;\beta}=& 0,\label{pe44}\\
    l_0^\beta m^\mu_{;\beta}=& \frac{1}{\omega} \left(l^\beta_{1;\alpha}- b^\beta_{;\alpha}\right)l_0^\alpha m_{ \beta} n^\mu \nonumber\\
    =& \frac{i}{\omega} R_{\alpha\beta\gamma\delta} l_0^\alpha m_0^\beta m_0^\gamma \tilde m_0^\delta n_0^\mu,\label{pe45}\\
    l_0^\beta \tilde m^\mu_{;\beta}=& \frac{1}{\omega} \left(l^\beta_{1;\alpha}- b^\beta_{;\alpha}\right)l_0^\alpha \tilde m_{ \beta} n^\mu \nonumber\\
    =& -\frac{i}{\omega} R_{\alpha\beta\gamma\delta} l_0^\alpha \tilde m_0^\beta \tilde m_0^\gamma m_0^\delta n_0^\mu.\label{pe46}
\end{align}
These equations assure that, up to the subleading order in $1/\omega$, the set of tetrads $\left(\dot x^\alpha, n^\alpha, m^\alpha, \tilde m^\alpha\right)$ satisfies the normalization and orthogonality relations in Eqs.~\eqref{oc18}-\eqref{oc20} throughout the ray. To verify this, we first simplify Eq.~\eqref{pe45} as
\begin{equation}
    l_0^\beta m^\mu_{;\beta} \approx \frac{1}{\omega} l_0^\beta m^\mu_{1;\beta}\approx \frac{1}{\omega} l_0^\beta \left(l_1^\alpha- b^\alpha\right)_{;\beta} m_{0\alpha} n_0^\mu.
\end{equation}
We can use the fact that the covariant derivatives of $m_{0\alpha}$ and $n_0^\mu$ are zero to write
\begin{equation}
    m_1^\mu= \left(l_1^\alpha-b^\alpha \right)m_{0 \alpha} n_0^\mu.\label{so48}
\end{equation}
We simplify Eqs.~\eqref{pe44} and \eqref{pe46} in a similar way to obtain
\begin{equation}
    n_1^\mu=0, \qquad \tilde m_1^\mu= \left(l_1^\alpha-b^\alpha \right) \tilde m_{0 \alpha} n_0^\mu,\label{so49}
\end{equation}
respectively. Noting that the Fermi propagated tetrad is constructed in Sec.~\ref{ntb} in such a way that it satisfies $w^\alpha l_{0\alpha}=0=w^\alpha n_{0\alpha}$, we can also write the subleading order of its component $\dot x^\alpha$ differently. Therefore, we can write $w^\alpha$ as
\begin{multline}
    w^\alpha \equiv \frac{1}{\omega} l_0^\beta \left(l_1^\alpha- b^\alpha\right)_{;\beta}=-\tilde \kappa m_0^\alpha- \kappa \tilde m_0^\alpha, \\ \kappa= - \frac{1}{\omega} m_0^\alpha l_0^\beta \left(l_{1\alpha}- b_\alpha\right)_{;\beta}.
\end{multline}
This allows one to write
\begin{equation}
    \left(l_1^\alpha- b^\alpha\right)= \tilde m_0^\beta \left(l_{1\beta}- b_\beta\right) m_0^\alpha + m_0^\beta \left(l_{1\beta}- b_\beta\right) \tilde m_0^\alpha. \label{so51}
\end{equation}
We can easily see that the subleading order correction of the tetrad $\left(\dot x^\alpha, n^\alpha, m^\alpha, \tilde m^\alpha\right)$, explicitly presented in Eqs.~\eqref{so48}, \eqref{so49} and \eqref{so51} satisfy the scalar products of Eqs.~\eqref{oc18}, \eqref{oc19} and \eqref{oc20}. The leading order terms of this tetrad are obviously the tetrad $\left(l_0^\alpha, n_0^\alpha, m_0^\alpha, \tilde m_0^\alpha\right)$ of Sec.~\ref{ntb}. Moreover, the subleading order terms of the Lorenz condition Eq.~\eqref{lc11} gives
\begin{equation}
    \frac{a_{;\mu}}{a}m_0^\mu=- m^\mu_{0; \mu}- i b_\mu m_0^\mu.\label{lc52}
\end{equation}

The field is not self-dual in the subleading order in $1/\omega$, since all the polarization Eqs.~\eqref{pe34}-\eqref{pe36} are not satisfied in the limit of spin optics (see Appendix~\ref{A2}).

\subsubsection{Self-dual solution up to the subleading order}

The tetrad $\left(\dot x^\alpha, n^\alpha, m^\alpha, \tilde m^\alpha\right)$, satisfying the scalar products of Eqs.~\eqref{oc18}, \eqref{oc19} and \eqref{oc20} is not the self-dual solution of the Maxwell equations. However, a self-dual solution of the Maxwell equations in Eqs.~\eqref{me4} and \eqref{me5} should exist for $J^\alpha=0$ in the subleading order geometric optics approximation. One can obtain this self-dual solution by first introducing the Fermi-like derivative operator
\begin{multline}
    {\cal D}'_l A^\alpha =l_0^\beta A^\alpha_{;\beta}-w_\beta A^\beta n^\alpha+ A^ \beta n_{ \beta} w^\alpha\\
    -\frac{i} {\omega} \left(\lambda_{;\mu}l^\mu m_\beta A^\beta m^\alpha- \tilde \lambda_{;\mu}l^\mu \tilde m_\beta A^\beta \tilde m^\alpha\right).
\end{multline}
The vanishing of the Fermi-like derivative ${\cal D}'_l A^\alpha=0$ gives
\begin{multline}
    l_0^\beta A^\alpha_{;\beta}= w_\beta A^\beta n^\alpha- A^ \beta n_{ \beta} w^\alpha\\
    +\frac{i} {\omega} \left(\lambda_{;\mu}l^\mu m_\beta A^\beta m^\alpha- \tilde \lambda_{;\mu}l^\mu \tilde m_\beta A^\beta \tilde m^\alpha\right),
\end{multline}
and this implies that the scalar product of any two tetrad components $\left(\dot x^\alpha, n^\alpha, m^\alpha, \tilde m^\alpha\right)$ is constant except that of $m^\alpha$ with itself and of $\tilde m^\alpha$ with itself. To see this, we calculate
\begin{multline}
    \left(a^\alpha b_\alpha\right)_{;\beta}l_0^\beta=a^\alpha b_{\alpha_;\beta}l_0^\beta+ b^\alpha a_{\alpha_;\beta}l_0^\beta\\
    = \frac{2 i} {\omega}\left( \lambda_{;\mu}l^\mu m_\beta a^\beta m^\alpha b_\alpha- \tilde \lambda_{;\mu}l^\mu \tilde m_\beta a^\beta \tilde m^\alpha b_\alpha\right).
\end{multline}
This scalar product is nonzero only if $a^\alpha =b^\alpha=m^\alpha$ or $a^\alpha =b^\alpha=\tilde m^\alpha$. For tetrads with vanishing Fermi-like derivatives, if they satisfy the following orthogonality and completeness relations at some point on the ray
\begin{align}
    & \dot x^\alpha m_{\alpha}= \dot x^\alpha \dot x_\alpha= \dot x^\alpha \tilde m_{\alpha}=0,  \qquad m^\alpha \tilde m_{ \alpha} =1,\\
    & n^\alpha m_{\alpha}= n^\alpha n_{\alpha}= n^\alpha \tilde m_{\alpha}=0, \qquad n^\alpha l_{\alpha}=-1,
\end{align}
then they satisfy these relations everywhere on the ray. However, $m^\alpha m_{\alpha}\ne 0$ and $\tilde m^\alpha \tilde m_{\alpha}\ne 0$, in general, along the circularly polarized ray in the subleading order approximation. This means that the polarization vectors are no more null as in the geometric optics limit. In addition to these, the tetrad evolves as
\begin{align}
    & l_0^\beta n^\alpha_{;\beta}= 0,\\
    & l_0^\beta m^\alpha_{;\beta}= w^\beta m_{ \beta} n^\alpha- \frac{i}{\omega} \tilde \lambda_{;\mu}l^\mu \tilde m^\alpha ,\label{pe62}\\
    & l_0^\beta \tilde m^\alpha_{;\beta}= w^\beta \tilde m_{ \beta} n^\alpha+ \frac{i}{\omega} \lambda_{;\mu}l^\mu m^\alpha.
\end{align}
As in Eqs.~\eqref{so48} and \eqref{so49}, we can write
\begin{align}
    &m_1^\mu= \left(l_1^\alpha-b^\alpha \right)m_{0 \alpha} n_0^\mu- i \tilde \lambda \tilde m_0^\mu, \qquad n_1^\mu=0, \label{pe64}\\
    &\tilde m_1^\mu= \left(l_1^\alpha-b^\alpha \right) \tilde m_{0 \alpha} n_0^\mu+ i \lambda m_0^\mu.
\end{align}
These tetrads components constitute the solution of the Maxwell equation in the Lorenz gauge, and they are self-dual, as they also satisfy the polarization Eqs.~\eqref{pe34}-\eqref{pe36}. Therefore, they provide a solution for the propagation of right-handed circularly polarized electromagnetic waves in curved spacetime in the spin optics approximation.

\subsubsection{Constructing a gauge independent Hamiltonian} \label{IVC5}

The gauge properties of the Berry connection and Berry curvature are related to the choice of the comoving frame. One can introduce noncanonical coordinates, such that the gauge-dependent Berry connection term~\cite[]{lf29} can be removed from the Hamiltonian of Eq.~\eqref{he65}. 
We consider the following transformation relations to the noncanonical coordinates
\begin{align}
    X^\alpha= & x^\alpha,\\
    L_\alpha= & l_{0\alpha}+ \frac{l_{1\alpha}-b_\alpha}{\omega}.
\end{align}
It is shown in~\cite{13} that such substitutions could be generated as the linearization of coordinates changes. 
The Hamiltonian Eq.~\eqref{he65} under this transformation becomes
\begin{align}
   H'(X,L)&= H(x,l)\\
   &= H\left(X^\alpha, L_\alpha- \frac{l_{1\alpha}-b_\alpha}{\omega}\right)\\
   &= H(X, L)- \frac{1}{\omega} \frac{\partial H_0}{\partial L_\alpha} (l_{1\alpha}-b_\alpha)\\
   &= H_0 (X,L),
\end{align}
where Eq.~\eqref{he65} is used in obtaining the last equality. Therefore, in the new coordinates $(X^\alpha, L_\alpha)$, the Hamiltonian reduces to
\begin{equation}
    H'(X,L)= \frac{1}{2} g^{\alpha\beta}(X) L_\alpha L_\beta.
\end{equation}
The corresponding Hamilton's equations of motion are
\begin{align}
    \begin{pmatrix}
    \dot X^\alpha\\
    \dot L_\alpha
    \end{pmatrix}= T' \begin{pmatrix}
    \frac{\partial H'}{\partial X^\beta}\\
    \frac{\partial H'}{\partial L_\beta}
    \end{pmatrix},
\end{align}
where $T'$ is the Poisson tensor in $(X,L)$~\cite[]{m34}. The Poisson tensor could be written as
\begin{align}
    T'= \begin{pmatrix}
    0  \quad & \delta^\alpha_\beta \\
    -\delta^\beta_\alpha \qquad  & \frac{k_{\alpha\beta}}{\omega}
    \end{pmatrix},
\end{align}
where $k_{\alpha\beta}$ is defined in Eq.~\eqref{k58} (it is compared with the Berry curvature term of condensed matter physics in Sec.~\ref{4c2}). Hamilton's equations of motion in the new variables are
\begin{align}
    \dot X^\alpha=& L^\alpha,\label{t90}\\ 
    \dot L_\alpha=& \Gamma^\mu_{\nu\alpha} L_\mu L^\nu+ \frac{1}{\omega} k_{\alpha\beta} L^\beta.
\end{align}
Covariant differentiation of Eq.~\eqref{t90} gives
\begin{equation}
    \frac{D \dot X^\alpha}{D\lambda}= \frac{D L^\alpha}{D\lambda}= \frac{1}{\omega} k^\alpha_\beta L^\beta,
\end{equation}
which is precisely the trajectory equation given in Eq.~\eqref{te43}. This shows that although the Hamiltonian of Eq.~\eqref{he65} contains the gauge-dependent term, the trajectory equation  obtained from it is gauge invariant. The reason is, as explained in Eq.~\eqref{gt11}, $U(1)$ gauge transformation freedom of the polarization basis $m^\alpha$ is constrained by the requirement that it should be Fermi propagated along the trajectory.

\section{Stress-energy tensor up to the subleading order} \label{s5}

To see how energy flows as waves propagate, we calculate the stress-energy tensor up to the subleading order approximation using the field tensor, given in Eq.~\eqref{ft32}. The stress-energy tensor due to the electromagnetic field is given by the relation
\begin{equation}
    4 \pi T_{\alpha\beta}= F_{\alpha\gamma} F_\beta^\gamma- \frac{1}{4} g_{\alpha_\beta} F_{\mu\nu} F^{\mu\nu}= \frac{1}{2} Re\left[ {\cal F}^{+1}_{\alpha\gamma} \tilde{\cal F}^{+1\gamma}_\beta\right],
\end{equation}
where $Re\left[ z\right]$ is the real part of $z$. This equation can be simplified by substituting $ {\cal F}^{+1}_{\alpha\gamma}$ from Eq.~\eqref{ft32}
\begin{equation}
    4 \pi T_{\alpha\beta}= \frac{\omega^2}{2} Re\left[{\cal Z}_{\alpha\gamma} \tilde{\cal Z}_\beta^\gamma \right].
\end{equation}
As ${\cal Z}_{\alpha\gamma}$ is self-dual, it can be expanded as (see Eq.~\eqref{pb31})
\begin{equation}
    {\cal Z}= \Phi_0 \mathbf{U}+ \Phi_1 \mathbf{W}+ \Phi_2\mathbf{V},
\end{equation}
where
\begin{align}
    \Phi_0=& \frac{1}{2}{\cal Z}_{\alpha\gamma} \left( l^\alpha m^\gamma- l^\gamma m^\alpha\right), \nonumber\\
    \Phi_2=&  \frac{1}{2}{\cal Z}_{\alpha\gamma} \left( \tilde m^\alpha n^\gamma- \tilde m^\gamma n^\alpha\right), \nonumber\\
    \Phi_1=&  -\frac{1}{4}{\cal Z}_{\alpha\gamma} \left( m^\alpha \tilde m^\gamma- m^\gamma \tilde m^\alpha- l^\alpha n^\gamma+ l^\gamma n^\alpha\right).
\end{align}
Substituting ${\cal Z}_{\alpha\gamma}$ from Eq.~\eqref{fa33}, we get
\begin{align}
    \Phi_0=& \frac{i a}{\omega} \sigma, \qquad \Phi_1= \frac{i a}{2 \omega} \left(\chi+ 2 \tau\right),\nonumber\\
    \Phi_2=& a- \frac{i a}{\omega}\left( -\frac{a_{,\alpha}}{a} n^\alpha- \tilde \mu+ \frac{\gamma- \tilde\gamma}{2}\right),
\end{align}
where $\sigma= - m^\alpha l_{\alpha;\beta} m^\beta$, $\chi= \tilde m^\beta m_{\beta;\alpha} m^\alpha$, $\tau= l^\beta m_{\beta;\alpha} n^\alpha$, $\mu= \tilde m^\beta n_{\beta;\alpha} m^\alpha$ and $\gamma- \tilde\gamma= -2 \tilde m^\alpha m_{\alpha;\beta} n^\beta$ are Newman-Penrose scalars. Thus, up to the subleading order in $1/\omega$,
\begin{multline}
    4 \pi T_{\alpha\beta}= \frac{\omega^2}{2} Re\Big[\Phi_0 \tilde\Phi_2 U_{\alpha \gamma}\tilde V_\beta^\gamma+ \Phi_1 \tilde\Phi_2 W_{\alpha \gamma}\tilde V_\beta^\gamma\\
    + \Phi_2 \tilde\Phi_0 V_{\alpha \gamma}\tilde U_\beta^\gamma+ \Phi_2 \tilde\Phi_1 V_{\alpha \gamma}\tilde W_\beta^\gamma + \Phi_2 \tilde\Phi_2 V_{\alpha \gamma}\tilde V_\beta^\gamma\Big].
\end{multline}
However, we have
\begin{align}
    & U_{\alpha \gamma}\tilde V_\beta^\gamma= \tilde m_\alpha \tilde m_\beta, \quad W_{\alpha \gamma}\tilde V_\beta^\gamma= -\tilde m_\alpha l_\beta- \tilde m_\beta l_\alpha, \nonumber \\
    & V_{\alpha \gamma}\tilde U_\beta^\gamma= m_\alpha m_\beta,\quad
    V_{\alpha \gamma}\tilde W_\beta^\gamma= -m_\alpha l_\beta- l_\alpha m_\beta, \nonumber \\
    & V_{\alpha \gamma}\tilde V_\beta^\gamma= l_\alpha l_\beta.
\end{align}
Collecting these values, we obtain, up to the subleading order in $1/\omega$,
\begin{align}
    4 \pi T_{\alpha\beta}&= \frac{\omega^2 a^2}{2} l_\alpha l_\beta+ \frac{i \omega a^2}{4}\bigg(2\left(\sigma \tilde m_\alpha \tilde m_\beta- \tilde \sigma m_\alpha m_\beta\right)\nonumber\\
    -& 2 \left(\mu-\tilde\mu+ \gamma- \tilde\gamma\right) l_\alpha l_\beta- \left(\chi+2 \tau\right)\left(\tilde m_\alpha l_\beta+ \tilde m_\beta l_\alpha\right) \nonumber\\
    & + \left(\tilde\chi+2 \tilde\tau\right)\left(m_\alpha l_\beta+ m_\beta l_\alpha\right) \bigg) \nonumber\\
    &= \frac{\omega^2 a^2}{2} L_\alpha L_\beta+ \frac{i \omega a^2}{2}\left(\sigma \tilde m_\alpha \tilde m_\beta- \tilde \sigma m_\alpha m_\beta\right),
\end{align}
where we redefined the wave vector as
\begin{multline}
    L_\alpha= l_\alpha+ \frac{i}{2 \omega}\Big(-\left(\mu-\tilde\mu+\gamma- \tilde\gamma\right) l_\alpha+ \left(\tilde\chi+2 \tilde\tau\right) m_\alpha\\
    - \left(\chi+2 \tau\right) \tilde m_\alpha \Big).
\end{multline}
The expression shows that the wave carries transverse stress due to the nonvanishing shear $\sigma$. The significance of this result is that we have obtained it without using the solution of the Hamilton-Jacobi equation~\eqref{hj47}. We could thus interpret this result as an independent verification of the spin Hall effect as depicted by the fact that the transverse stress of the wave is frequency-dependent. The energy-momentum tensor is not unique, in the limit of spin optics, in the sense that the tensor
\begin{equation}
    \Theta^{\mu\nu}= T^{\mu\nu}+ \nabla_\lambda A^{\mu\nu\lambda}; \qquad A^{\mu\nu\lambda}= -A^{\mu\lambda\nu}
\end{equation}
also satisfies the criterion to be the energy-momentum tensor. Thus, it is not a surprise that with the appropriate choice of $A^{\mu\nu\lambda}$ we could reduce this energy-momentum tensor into the expression given in Ref.~\cite{d36}, which is derived by a similar mathematical formulation. As expected, the direction of energy flow is not along the direction of the wave vector $l^\alpha$, a characteristic of waves propagating in anisotropic medium.

\section{Comparison of results} \label{s6}

Ref.~\cite{13} also developed a covariant formulation of the spin Hall effect. Our results differ slightly from theirs, primarily owing to the difference in our eikonal function Eq.~\eqref{a8} with theirs. In their formulation, the amplitude $a^\alpha$ is assumed to be a function of the phase gradient $l^\alpha$, that is, $a^\alpha= a^\alpha \left(\lambda, l(\lambda) \right)$.
Using the WKB analysis with this form of eikonal function, the following equation for the ray trajectory was obtained:
\begin{equation}
    \dot x^\gamma= \frac{1}{\omega}\left(l^\gamma -B^\gamma- l_\mu \frac{\partial B^\mu}{\partial l_\gamma}\right),
\end{equation}
where
\begin{align}
    &B_\beta (\lambda, l(\lambda))= \frac{i}{2}\left(\tilde m^\alpha \overset{h}{\nabla}_\beta m_{\alpha}- m^\alpha \overset{h}{\nabla}_\beta \tilde m_{\alpha} \right)= i \tilde m^\alpha \overset{h}{\nabla}_\beta m_{\alpha}, \nonumber\\
    &\overset{h}{\nabla}_\beta m_{\alpha}= \nabla_\beta m_{\alpha}+ \Gamma^\mu_{\beta\nu} l_\mu \frac{\partial m_\alpha}{\partial l_\nu}.
\end{align}
Thus, if we start with the WKB expansion whose amplitude does not depend on the phase gradient, that is, if we take
\begin{equation}
    a^\alpha (\lambda, l(\lambda)) = a^\alpha (\lambda). \label{a94}
\end{equation}
then we could show that the ray trajectory derived by Ref.~\cite{13} would be the same as the propagation equation of Eq.~\eqref{te43}.

Ref.~\cite{14} started with the following form of the WKB expansion for the vector potential:
\begin{equation}
    A^\alpha=a^\alpha(\lambda) e^{i {\cal S}},
\end{equation}
where $l_\alpha={\cal S}_{;\alpha} /\omega$ is the wave vector, and $\lambda$ is the usual parameter from above. The only difference in this eikonal function from the one given in Eq.~\eqref{a8} is: here, the phase gradient is not expanded in powers of $1/\omega$, that is, $l^\alpha=l_0^\alpha$. Thus, all of our Eqs.~\eqref{dr50}-\eqref{lc52} for spin optics would be the same as Frolov's if $l_1^\alpha=0$. However, as explained above Eq.~\eqref{gt11}, we could not take $l_1^\alpha=0$ as we are using parallel propagated tetrad in the leading order approximation.

\section{Discussion and conclusions} \label{s7}

We formulated the theory of spin optics using WKB formalism, where we expand both the amplitude and phase in powers of $1/\omega$, $\omega$ being the characteristic frequency. This expansion in both the amplitude and phase is necessary if we want to simplify the problem by imposing gauge conditions (for example, we needed to use the Fermi propagated tetrad here) that reduce some of the freedom in the transformation of the null tetrad. This novel application of the WKB formalism is used to derive the evolution equations for both the propagation vector and polarization tensor. We are aware of only the evolution equation for the propagation vector derived in the literature. In the geometric optics approximation, polarization and propagation vectors are two of the components of a parallel propagated null tetrad. This requirement of parallel propagation, which is necessary in the subleading order also, restricts some of the freedom in the transformation of the null tetrad, thereby resulting in the observer/ gauge independent spin Hall effect. We have proved that our result is indeed observer independent in Sec.~\ref{IVC5} and thus distinguishes itself from the observer/emitter-dependent effects occurring in the geometric optics regime~\cite{dt51}. Such effects are:
1) the special relativistic effect due to the change in polarization direction occurring from the Wigner rotation and 
2) the general relativistic effect due to the absence of a global reference direction, making us unable to determine a unique standard polarization triad in curved spacetime.
Moreover, we have pointed out in Ref.~\cite{dt52} that the initially divergent trajectories do not reconverge; the separation between the rays of different frequencies increases with increasing distance. We could draw the trajectories from numerical calculations, confirming that the initially divergent trajectories do not reconverge. This is because the gravitational spin Hall effect results from spin-orbit interaction, and neither spin nor orbital angular momentum reverses in direction as the particle crosses the distance of the closest approach, causing the effect of spin-orbit interaction to increase, not decrease.

One of the ways to understand the nature of light propagation in curved spacetime is to carry forward an analogy from condensed matter physics, where the phenomena are well understood and experimentally verified~\cite[]{5,1,e18,e19}. Notably, this analogy is beneficial when several authors have approached this problem with the eikonal formalism itself and obtained slightly different results. The analogy with the optical Magnus effect from condensed matter physics revealed that the Berry phase and spin Hall effect are closely related to the dynamics of the intrinsic angular momentum of the  wave. In particular, the spin Hall effect results from the bending of the trajectory of photons with nonzero spin. Moreover, the transverse deflection of the ray trajectory is proportional to the curvature of the ray~\cite[]{5}(or equivalently to the curvature of the spacetime, where the ray is propagating~Eq.\eqref{te43}).


The derivation of the spin Hall effect made here is based on classical arguments. However, one can make the quantum mechanical interpretation in the following manner. The geometric optics approximation is valid in the infinite frequency limit, which implies the evolution of the confined wave packet. The spin Hall effect arises while extending this approximation to waves of finite but large frequency. This effect is due to the interference of multiple partial plane waves constituting the wave packet that propagates in slightly different directions and thus acquires slightly different geometric phases. Therefore, because of the transverse gradient of the Berry phase, which is the phase of the plane wave in the packet, the spin Hall effect is observed.

\acknowledgments

PKD is supported by an International Macquarie University Research Excellence Scholarship.









\appendix

\section{Geometric optics limit} 

All the results of geometrical optics could be retrieved by taking Eqs.~\eqref{lc11}-\eqref{dr13} and substituting $m_1^\beta=0=l_1^\beta$. The Lorenz condition Eq.~\eqref{lc11} and the wave equation Eq.~\eqref{we12} in the leading order approximation in $\omega$ reduces to
\begin{equation}
    l_0^\alpha m_{0 \alpha}= 0= l_0^\alpha l_{0\alpha}.\label{go15}
\end{equation}
Next, we calculate $\tilde m_{0\alpha}j^\alpha$ from Eq.~\eqref{we12} by taking $m_1^{\alpha}=0=l_{1\mu}$ as they are subleading order terms in $\omega$ and thus irrelevant in geometric optics approximation, to obtain
\begin{equation}
    l^\beta_{0;\beta}+2 \tilde m_{0 \alpha} m^\alpha_{0;\beta}l_0^\beta+2 \frac{a_{;\beta}} {a} l_0^\beta=0.
\end{equation}
Since the term $\tilde m_{0 \alpha} m^\alpha_{0;\beta}l_0^\beta$ is purely imaginary and the remaining terms
\begin{equation}
    l^\beta_{0;\beta}+2 \frac{a_{;\beta}} {a} l_0^\beta, \nonumber
\end{equation}
are purely real, they should be separately zero, thereby giving
\begin{equation}
    l^\beta_{0;\beta}+2 \frac{a_{;\beta}} {a} l_0^\beta=0, \qquad  m^\alpha_{0;\beta}l_0^\beta=0. \label{go17}
\end{equation}
In the geometric optics approximation, these are the entire set of equations for electromagnetic waves in curved spacetime.

\section{Checking self-duality} \label{A2}

To verify that the tetrad $\left(\dot x^\alpha, n^\alpha, m^\alpha, \tilde m^\alpha\right)$, satisfying Eqs.~\eqref{so48}, \eqref{so49} and \eqref{so51} in the subleading order, is not a self-dual solution, let us first calculate Eq.~\eqref{pe36}
\begin{multline}
    {\cal Z}_{\alpha\beta} l^\alpha \tilde m^\beta=\frac{i a}{\omega} \left(-\frac{a_{;\alpha}}{a}l_0^\alpha+m_{0\alpha;\beta}l_0^\alpha \tilde m_0^\beta\right)\\
    =\frac{i a}{\omega} \left(\frac{1}{2} l^\alpha_{0;\alpha}-m_0^\alpha l_{0 \alpha;\beta} \tilde m_0^\beta\right)=0,
\end{multline}
where Eq.~\eqref{go17} is used to obtain this identity. Similarly, Eq.\eqref{pe35} gives
\begin{multline}
    {\cal Z}_{\alpha\beta} \left(\tilde m^\alpha m^\beta- l^\alpha n^\beta \right)=\frac{i}{\omega}\left(\frac{a_{;\alpha}}{a}m_0^\alpha-m_{0\alpha;\beta}l_0^\alpha n_0^\beta\right)\\
    =\frac{i}{\omega}\left(- m^\alpha_{0; \alpha}- i b_\alpha m_0^\alpha-m_{0\alpha;\beta}l_0^\alpha n_0^\beta\right)=0,
\end{multline}
where Eq.~\eqref{lc52} is used to arrive at this identity. Finally, Eq.~\eqref{pe34} gives
\begin{multline}
    {\cal Z}_{\alpha\beta} m^\alpha n^\beta=\frac{i}{\omega}\left(-m_{0\alpha;\beta}m_0^\beta n_0^\alpha\right)\\
    =\frac{i} {\omega}\left(m_0^\alpha n_{0\alpha;\beta} m_0^\beta\right)\equiv \frac{i} {\omega} \tilde \lambda,
\end{multline}
where $\lambda$ denotes the Newman-Penrose scalar. Hence Eq.~\eqref{pe34} is not satisfied unless $\tilde \lambda=0$.

\end{document}